\documentclass{emulateapj}

\begin{document}

\def\fluxthres{\hat f_{\bar \e}}
\def\fluxeps{f_{_{\rm \epsilon}}}
\def\zmin{z_{\rm min}}
\def\zmax{z_{\rm max}}
\def\xmin{x_{\rm min}}
\def\xmax{x_{\rm max}}
\def\e{\epsilon}
\def\Estar{{\cal E}_*}
\def\Estarg{{\cal E}_{*\gamma}}
\def\Estargo{{\cal E}_{*\gamma 0}}
\def\Swift{\emph{Swift}}

\newcommand{\begeq}{\begin{equation}}
\newcommand{\fineq}{\end{equation}}
\newcommand{\begfig}{\begin{figure}}
\newcommand{\finfig}{\end{figure}}
\newcommand{\begeqarray}{\begin{eqnarray}}
\newcommand{\fineqarray}{\end{eqnarray}}

\slugcomment{Accepted for publication in \apj}

\shorttitle{GRB Predictions for the Fermi Telescope} \shortauthors{Le \& Dermer}

\title{Gamma Ray Burst Predictions for the Fermi Gamma Ray Space Telescope}

\author{Truong Le\altaffilmark{1} and Charles D. Dermer\altaffilmark{2}}
\affil{Space Science Division, Code 7653 \break Naval
Research Laboratory,  Washington, DC 20375, USA}

\altaffiltext{1}{tle@ssd1.nrl.navy.mil; now at Space Telescope Science Institute, Baltimore, MD}
\altaffiltext{2}{charles.dermer@nrl.navy.mil}

\begin{abstract}
Results of a phenomenological model to estimate the GRB detection
rate by the Fermi Gamma ray Space Telescope are reported. This
estimate is based on the BATSE 4B GRB fluence distribution, the
mean ratio of fluences measured at $100$ MeV -- 5 GeV with EGRET
and at $20$ keV -- 2 MeV with BATSE, and the mean EGRET GRB
spectrum for the 5 EGRET spark-chamber GRBs. For a 10\% fluence
ratio and a number spectral index $\alpha_1 = -2$ at 100 MeV -- 5
GeV energies, we estimate a rate of $\approx 20$ and 4 GRBs per yr
in the Fermi Large Area Telescope field of view with at least 5
photons with energy $E > 100$ MeV and $E > 1$ GeV, respectively.
We also estimate $\approx 1.5$ GRBs per yr in the Fermi FoV where
at least 1 photon with energy $E > 10$ GeV is detected.  For these
parameters, we estimate $\approx 1$ -- 2 GRBs per year detected
with the Fermi telescope with more than 100 $\gamma$ rays with
$E\gtrsim 100$ MeV. Comparison predictions for $\alpha_1 = -2.2$,
different fluence ratios, and the AGILE $\gamma$-ray satellite are
made. Searches for different classes of GRBs using a diagram
plotting $100$ MeV -- 10 GeV fluence vs.\ 20 keV -- 20 MeV fluence
is considered as a way to search for separate classes of GRBs and,
specifically, spectral differences between the short-hard and long
duration GRB classes, and for hard components in GRBs.
\end{abstract}


\section{Introduction}

The Gamma ray Large Area Space Telescope, GLAST, was launched on
11 June 2008 and renamed the Fermi Gamma ray Space Telescope
(hereafter Fermi) on  August 26, 2008. Fermi comprises two
separate instruments, the Large Area Telescope (LAT), which is
sensitive to $\gamma$ rays in the energy range from $E\approx 30$
MeV to $ E >$ 300 GeV, and the Fermi GLAST Burst Monitor (GBM),
sensitive at 10 keV $\lesssim E \lesssim$ 30 MeV. At GeV energies,
the Fermi LAT provides an increase in effective area over the
Energetic Gamma Ray Experiment Telescope (EGRET) on the Compton
Gamma Ray Observatory (CGRO) by almost an order of magnitude, a
smaller point spread function, a field-of-view (FOV) larger by a
factor $\approx 5$ than EGRET, and the capability to autonomously
slew the LAT in response to a trigger from the GBM. The GBM uses
NaI scintillators for triggering, like the Burst and Transient
Source Experiment (BATSE), in addition to BGO (Bismuth Germanate)
detectors for response in the $\sim 0.2$ MeV -- 25 MeV energy
range. The GBM views the unocculted sky, $\approx 50$\% of the
full sky, and the predicted long-duration GBM burst detection rate
is $\approx 150$ -– 225 per year when triggering using standard
BATSE detection criteria \citep[$>4.5\sigma $ in at least two
detectors in 1.024 s in the 50 –- 300 keV;][]{kip01}. This can be
compared with the BATSE detection rate of $\approx 550$ GRBs
full-sky brighter than $0.3$ cm$^{-2}$ s$^{-1}$ in the 50 -- 300
keV band \citep{ban02}.

Five EGRET $\gamma$-ray transients, coincident in time and
direction with a BATSE GRB, were detected with the EGRET spark
chamber at $E> 30$ MeV with photon counts in excess of background
\citep{din95}. Knowledge from the CGRO can be used to quantify
expectations for Fermi. To simulate GRBs in the Fermi energy band,
we calculate the average fluence ratio between EGRET (100 MeV -- 5
GeV) and BATSE ($\simeq 20$ keV -- $2$ MeV) from the 5 BATSE GRBs
that were detected with the EGRET spark chamber. Using an average
spectrum consisting of an empirical \citet{ban93} spectrum plus a
power law, and the effective area of the Fermi LAT, we estimate
the full-sky number of GRBs per year with different numbers of
high-energy photons based on the BATSE 4B fluence distribution for
long-duration GRBs \citep{pac99}. We make predictions for both
long-duration and total GRBs; only the former class of GRBs were
convincingly detected at 100 MeV -- 5 GeV energies with CGRO. Both
Fermi and AGILE view $\approx 1/5^{th}$ of the full sky at 100 MeV
-- 5 GeV energies, with the field-of-view defined by the solid
angle where the effective area is $\gtrsim$ one-half the on-axis
effective area. Thus we consider the realistic detection rate to
be $\approx 0.2\times$ the full-sky rate, without considering the
potential effects of autonomous repointings in the estimates.

In \S~2 we discuss the phenomenological model for GRBs based on
BATSE and EGRET data, and estimate the uncertainty in GRB
properties derived from EGRET and Fermi measurements. The
predicted GRB detection rates and GRB contribution to the diffuse
extragalactic background radiation are calculated in \S~3. We note
that these predictions were made in advance of the launch of Fermi
(see arXiv:0807.0355 v1), so can be used to test our knowledge of
GRB physics, in particular, whether additional hard, high-energy
($\gtrsim 30$ MeV) $\gamma$-ray spectral components are present in
GRBs.

\section{Model}

We develop a phenomenological model of GRBs based on the
properties of EGRET spark-chamber GRBs, all detected with BATSE,
and then make predictions for the GRB detection rate with Fermi
and AGILE. Based on BATSE and EGRET results, and from earlier GRB
studies, the time-integrated (and resolved, where possible) GRB
photon number spectrum $N(E_\gamma )$ is assumed to be
well-described by the sum of the Band function \citep{ban93},
$N_{\rm B}(\e )$, and a high-energy power law component,
$N_G(\epsilon )$. We write this function using dimensionless
notation $\epsilon = E_\gamma / m_e c^2$ as
\begin{equation}
N(\epsilon ) = N_{\rm B}(\epsilon ) + N_G(\epsilon )
\label{nepsilon}
\end{equation}
where $N_{\rm B}(\e)$ takes the form
\begin{eqnarray}
N_{\rm B}(\epsilon) & = & k_{_B} \; \epsilon^{\alpha}
\exp[-\epsilon (\alpha-\beta)/\epsilon_{_{\rm br}}] \; H(\epsilon;
\e^{\rm B}_{\rm min},\epsilon_{_{\rm br}}) \nonumber \\
 & + & \;  k_{_B} \; \epsilon^\beta \; \epsilon_{_{\rm br}}^{\alpha-\beta}
\exp(\beta-\alpha)
\; H(\e;\e_{\rm br},\e^{\rm B}_{\rm max}) \; .  %
\label{Nband-eq} %
\end{eqnarray}
The parameters $\alpha$ and $\beta$ are the low- and high-energy
Band $\alpha$ and Band $\beta$ indices, and $E_{\rm br} =
m_ec^2\epsilon_{_{\rm br}}\sim 100$ keV is the ``break energy.''
The photon energy $E_{\rm pk}$ of the peak of the $\nu F_\nu$
spectrum is related to $E_{\rm br}$ through the expression
\begin{equation}
E_{\rm pk} = {(2+\alpha)E_{\rm br}\over \alpha - \beta }\;,
\label{Epk}
\end{equation}
provided $\alpha >-2$ and $\beta <-2$. The Heaviside function is
defined such that $H(x;y,z)=0$ except at $y\leq x\leq z$, where
$H(x;y,z)=1$ . The term $k_{_B}$ is the constant normalizing the
number fluence to the $> 20$ keV BATSE energy fluence $\Phi_{_B}
(> 20 {\rm \; keV})$ of a particular GRB, and is given by $k_{_B}
= \Phi_{_B} (> 20 {\rm \; keV})/I_{_1}$, where
\begin{eqnarray}
{I_{_1}\over m_e c^2} & = &  \int^{\epsilon_{_{\rm
br}}}_{\epsilon^{\rm B}_{_{\rm min}}} \epsilon^{\alpha + 1}
e^{-\epsilon (\alpha - \beta)/\epsilon_{_{\rm br}}} \; d\epsilon\ \nonumber \\
& + & \; \epsilon_{_{\rm br}}^{\alpha - \beta} e^{\beta - \alpha}
\int^{\epsilon^{\rm B}_{_{\rm max}}}_{\epsilon_{_{\rm
br}}} \epsilon^{\beta + 1} \; d\epsilon  \; . %
\label{I1-eq} %
\end{eqnarray}

For model GRBs normalized to the BATSE fluence, we take the
minimum energy $\epsilon^{\rm B}_{_{\rm min}}=30$ keV, and maximum
energy $\epsilon^{\rm B}_{_{\rm max}}= 5$ GeV. For the mean GRB
spectral indices, we take $\alpha = -0.9$, $\beta = -2.2$, and
break energy $E_{_{\rm br}}= 250$ keV, corresponding to the mean
for these values from spectral analyses of bright BATSE GRBs
\citep{pre00}.

For the high-energy spectral component, we assume that GRBs can be
described by a power law at $\gamma$-ray energies written as
\begin{equation}
N_{_G}(\epsilon) = k_{_G} \epsilon^{\alpha_{_1}} H(\epsilon; \;
\epsilon_{_{\rm min}}, \epsilon_{_{\rm max}}) \; .%
\label{NG-eq} %
\end{equation}
The average spectrum of GRBs measured by the EGRET spark chamber
is $\alpha_1 = -1.95\pm 0.25$ for 45 $\gamma$ rays detected within
200 seconds of the trigger \citep{din95,din98}, similar to the
average EGRET Total Absorption Shower Counter (TASC) index
\citep{cds98}. Band $\beta$ values for BATSE GRBs have a large
range, with most GRBs showing  $ -2 \lesssim \beta \lesssim 2.5$,
but lie in a different range than EGRET and Fermi emissions. The
mean spectral index measured by EGRET for the five spark-chamber
GRBs is not in good agreement with the index of BATSE and TASC
GRBs at $E\gtrsim 1$ MeV \citep{pre00}. This suggests that the
spectrum is probably not a simple power law at high energies, but
different spectral shapes should be considered \citep[e.g.,
cut-offs or extra components; see][]{kgp08}, and unusual spectral
evolution of high-energy components \citep{gon03}.

Since the spectral index distribution of GRBs at medium
$\gamma$-ray energies is still debatable, which Fermi will
clarify, in this paper we make predictions assuming that the
$>100$ MeV $\gamma$-ray spectrum can be described by a simple
power law with $\alpha_1 \cong -2$. For definiteness, we consider
$\alpha_1 = -2$ and $\alpha_1 = -2.2$. The latter value
corresponds to the softest spectrum within 1$\sigma$ of the EGRET
measurements, and is the same as the Band $\beta$ used,
effectively giving a single power law without break. Besides
spectral index, our predictions are also strongly affected by
EGRET uncertainties, which we will discuss in the next section.

\begin{figure}[t]
\includegraphics[width=3.0in]{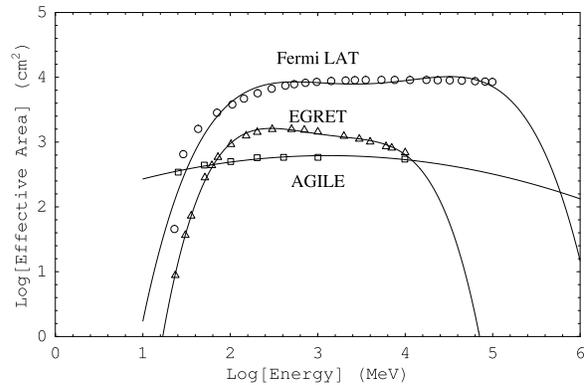}
\caption{Approximate on-axis effective area of the Fermi LAT, EGRET, and AGILE.
The symbols are the data, and
the solid lines are the best-fitted functions.\\} %
\label{fig1} %
\end{figure}

The term
\begin{equation}
k_{_G}= {\rho \Phi_{_B}(> 20
{\rm~keV})\over I_{2}}\;
\label{kG}
\end{equation}
where
\begin{equation}
{I_{2}\over m_e c^2} = \int^{\epsilon{_{\rm max}}}_{\epsilon_{_{\rm
min}}} \;d\e \;\epsilon^{\alpha_{_1} + 1}
\label{I_2}
\end{equation}
is the normalization constant based on the average fluence ratio
$\rho$ between EGRET and BATSE. In our model, the uncertainties in
the number of $\gamma$ rays that were detected by EGRET and the
measured BATSE and EGRET fluences are absorbed into this factor
$\rho$. Here $\epsilon_{_{\rm min}}$ and $\epsilon_{_{\rm max}}$
are the minimum and maximum energy at 100 MeV and 5 GeV,
respectively; note that even though EGRET detected photons down to
30 MeV energies, the effective area dropped rapidly below $\approx
100$ MeV. The higher energy given by $\epsilon_{_{\rm max}}$ is
set by the energy where self-vetoing in EGRET reduces its
effective area (see Fig.~\ref{fig1}).

The number of source counts with energy $\epsilon >
\epsilon_{\gamma}$ for different detectors \citep{dd04} is given
by
\begin{equation}
S_i(>\epsilon_\gamma) \cong f_{_\gamma}
\int^{\infty}_{\epsilon_{\gamma}} A_i(\epsilon) \;
N(\epsilon) \; d\epsilon \; , %
\label{Source-eq} %
\end{equation}
where $A_i(\e )$ is the on-axis effective area of different
detectors (e.g., EGRET, Fermi LAT, and AGILE) and $f_{_\gamma}$ is
a collection function depending on the opening angle of the
detector. Here we conservatively take the opening angle as
energy-dependent with $f_{_\gamma} = 68\%$ containment
\citep{tho86,dd04} though a larger region of interest (ROI) will
not much change the results for bright GRBs with many $\gamma$
rays. The on-axis effective areas for EGRET ($A_E$), Fermi LAT
($A_F$), and AGILE ($A_A$) are taken from the best fit to the
calibration and analysis data \citep[][and references
therein]{pit04,bel07} as shown in Fig.\ \ref{fig1}. The functions
describing these areas are
\begin{eqnarray}
A_E(x) & = & 10^{a_0 + a_1 x + a_2 x^2 + a_3 x^3 + a_4 x^4} \nonumber \\ %
A_F(x) & = & 10^{a_0 + a_1 x + a_2 x^2 + a_3 x^3 + a_4 x^4}  \\
A_A(x) & = & 10^{a_0 + a_1 x + a_2 x^2} \; , \nonumber %
\label{Earea-eq} %
\end{eqnarray}
where $x=\log E$, and parameters are given in Table~1. Most GRBs
would take place at relatively large angles, $\approx 30$ --
60$^\circ$, from the Fermi axis, where the effective area is half
this value. On the other hand, a larger ROI would include more
counts. These represent two additional uncertainties in the rate
estimates.

\begin{deluxetable}{cccccc}
\tabletypesize{\small} %
\tablecaption{Effective Area Parameters\label{tbl-1}} %
\tablewidth{-0.pt} %
\tablehead{ %
\colhead{$\rm Effective \atop {Area \atop (cm^2)}$} %
&\colhead{$a_0$} %
&\colhead{$a_1$} %
&\colhead{$a_2$} %
&\colhead{$a_3$} %
&\colhead{$a_4$} %
} \startdata
$A_E$ %
& $-24.116 $   %
& $36.435  $   %
& $-18.000 $   %
& $3.912   $   %
& $-0.317  $   %
\\
$A_A$ %
& $2.010  $    %
& $0.501  $    %
& $-0.080 $    %
\\
$A_F$ %
& $-11.583   $   %
& $18.503    $   %
& $-8.127    $   %
& $1.555     $   %
& $-0.109    $   %
\enddata
\end{deluxetable}

The on-axis effective area of Fermi is shown in Fig.\ \ref{fig1}.
Signal analysis of photon events in the Fermi LAT depends on event
reconstruction and background rejection \citep{atw09}. The {\it
diffuse} class of events, for instance, minimizes background by
rejecting tracks without good energy measurement in the
calorimeter and requires both tracker and calibration information.
GRBs can, however, be studied with the {\it transient} event
class, which increases low-energy response at the expense of
greater background using large regions of interest around the GRB
direction. We consider only photons with  $E > 100$ MeV, where the
point spread function and consequently the background is smaller
than for photons at lower energies, limited at higher energies to
a maximum of 5 GeV. Over this energy range, both the EGRET and
Fermi effective area is relatively flat, and measurement
uncertainty can be more reliably estimated.

\begin{deluxetable*}{lccccccccccc}
\hskip-1.in
\tabletypesize{\scriptsize} %
\tablecaption{EGRET Gamma-Ray Burst Spark Chamber Detections\label{tbl-2}} %
\tablewidth{0pt} %
\tablehead{ %
\colhead{} & %
\colhead{} & %
\colhead{} & %
\colhead{} &%
\colhead{} & %
\colhead{} & %
\multicolumn{2}{c}{EGRET GRB $\gamma$ rays}    & %
\multicolumn{2}{c}{spectral index of $-2$}    & %
\multicolumn{2}{c}{spectral index of $-2.2$}  \\ %
\colhead{$\rm Burst \atop Date$}       & %
\colhead{$\theta^a$}  & %
\colhead{$N_{sc}^b$}  & %
\colhead{$N_{>100}^c$} & %
\colhead{$E_{max}({\rm GeV})$} & %
\colhead{$\Phi^d_{_{\rm BATSE}} $}  & %
\colhead{$\Phi^e_{_{\rm EGRET}}  $} & %
\colhead{$\rho(\% )^f$}   & %
\colhead{$\Phi^e_{_{\rm EGRET}}  $} & %
\colhead{$\rho(\% )^f$}   & %
\colhead{$\Phi^e_{_{\rm EGRET}}  $} & %
\colhead{$\rho(\%)^f$}    %
} %
\startdata
910503    %
& 24      %
& 9$^g$   %
& 2       %
& 10      %
& $2.92(\pm 0.01) $ %
& $1.28$ %
& $0.43$ %
& $2.68$ %
& $0.92$ %
& $2.20$ %
& $0.75$
\\
910601 %
& 12   %
& 8    %
& 3    %
& 0.31 %
& $1.5 (\pm 0.01)$ %
& $1.13$ %
& $0.75$ %
& $1.79$ %
& $1.19$ %
& $1.48$ %
& $0.99$
\\
930131 %
& 28    %
& 18    %
& 12    %
& 1.2    %
& $0.66(\pm 0.11)$ %
& $26.6$ %
& $40.3$ %
& $24.8$ %
& $37.6$ %
& $20.3$ %
& $30.8$
\\
940217 %
& 10    %
& 28    %
& 10    %
& 18    %
& $6.6 (\pm 0.03) $ %
& $13.0$%
& $1.97$ %
& $5.53$ %
& $0.84$ %
& $4.58$ %
& $0.69$
\\
940301 %
& 6    %
& 7    %
& 5    %
& 0.16    %
& $1.12(\pm 0.01)  $ %
& $1.14$  %
& $1.02$   %
& $2.48$  %
& $2.21$  %
& $2.05$  %
& $1.83$
\enddata
\tablecomments{
\noindent $^a$Angle from axis in degrees\\
\noindent $^b$Number of $\gamma$ rays imaged by EGRET spark chamber\\
\noindent $^c$Number of $\gamma$ rays imaged by EGRET spark
chamber between 100 MeV and 5 GeV \\
\noindent $^d$BATSE 20 keV -- 2 MeV fluence, units of $10^{-4}$
erg cm$^{-2}$, from \citet{mee96}\\
\noindent $^e$EGRET 100 MeV -- 5 GeV fluence, units of $10^{-6}$ erg cm$^{-2}$\\
\noindent $^f\rho={\Phi_{\rm EGRET}}/{\Phi_{\rm BATSE}} $\\
\noindent $^g$Nine events within 15 s after GRB trigger
\citep{sch92}}
\end{deluxetable*}

\section{Fluence Estimates and Uncertainties for EGRET GRBs}

In Table~2, we list the pertinent data from EGRET to make
estimates of Fermi detection rates. The first five columns of
Table~2 give the GRB date-name, the angle of the GRB with respect
to the z-axis of the EGRET telescope, the total number and the
number of $\gamma$ rays with energies between 100 MeV and 5 GeV
imaged by the EGRET spark chamber, and the maximum $\gamma$-ray
energy associated with the GRB. With this data, we can calculate
the 100 MeV  -- 5 GeV fluence $F_{0.1{\rm-}5{\rm~ GeV}}$ from an
EGRET spark chamber GRB using the expression
\begin{equation}
F_{0.1-5{\rm~ GeV}}= \sum_i{E_i\over A(E_i,\theta)}\;,
\label{fluence}
\end{equation}
with the sum being over the $i = 1,\dots, N_\gamma$ $\gamma$ rays
with energies from 100 MeV to 5 GeV detected by the EGRET spark
chamber and associated with the GRB, located at the angle $\theta$
with respect to the pointing axis of the EGRET.

By choosing a specific energy window between 100 MeV and 5 GeV, we
can minimize uncertainty in the GRB fluence estimate. This is
where the EGRET response is best understood, as compared to the
higher energy regime with uncertain systematics \citep{shk08}, or
at lower energies where the point spread function increases and
the numbers of background photons are larger. Because the
background photons are more likely to be at low energies, for
example, due to difficulties in event reconstruction, the number
of background photons, quoted below from the analysis of each
specific EGRET GRB, is an upper limit to the number of background
photons in the 100 MeV -- 5 GeV range.

\subsection{Fluence Error Estimate for EGRET Spark Chamber GRBs}

Several errors come into the estimate of the 100 MeV -- 5 GeV
burst fluence, including (i) the statistical error due to the
finite number of detected photons; (ii) deadtime effects in the
EGRET detector; and (iii) background spark-chamber events
erroneously associated with the GRB. Other effects may also play a
role, including (iv) errors in energy resolution ($\approx 12$ \%
and 9\% $\gamma$-ray photon energy resolution errors at 100 MeV
and 1 GeV, respectively), and (v) errors in effective area. For
the EGRET effective area, we use the angle-dependent EGRET
effective area with the TASC in coincidence \citep[Fig.\ 14
of][]{tho93}, where the effective area for the associate burst
(See Table~\ref{tbl-2}) is obtained by interpolating the data in
Figure~14.

For GRB 910503, 9 photons are detected within 15 s of the GRB
trigger, and 6 are likely to be associated with this GRB, implying
a possible background of 3 events \citep{sch92}. A 10 GeV $\gamma$
ray was also detected 84 s after this GRB \citep{mer95} coincident
with its location. Six of the $\gamma$ rays were detected within 2
seconds, allowing as much as a factor of $\approx 30$\% error for
0.1 s EGRET deadtime if high-energy $\gamma$ rays are produced
roughly uniformly over this time interval. If such emission is
produced in bursts shorter than the $\sim 0.1$ s deadtime, then
the GRB produced a stronger fluence then measured. For GRB 910601
\citep{kwo93}, 4 events were detected within 100 s of the GRB
trigger, with 3 events greater $100$ MeV $\gamma$ rays, when only
1.5 events were expected.  This represents a significant
uncertainty on the fluence due to background. In contrast,
deadtime effects are unlikely to be important for GRB910601
assuming, as we do henceforth, that $\gamma$-rays are not produced
in sub-100 ms bursts.

GRB 930131, the so-called ``superbowl burst," is interesting in
that it shows an intense $\sim 0.1$ s peak characteristic of a
short hard GRB \citep{som94}, followed $\approx 0.8$ s later by a
weaker $\approx 0.4$ s pulse, and also displays extended emission
detected with BATE for ~50 s after the GRB trigger \citep{kou94}.
Although its $t_{90}$(50 -- 300 keV) $= 19.200\pm 2.56$ s, its
$t_{50}$(50 -- 300 keV) $ = 1.024\pm 0.091$ s \citep{pac99}, so it
is near the boundary between the long soft and short hard GRBs
\citep{kou93}. Note that short hard GRBs are often observed with
Swift to produce extended emission \citep{ng08}. Sixteen $\gamma$
rays are imaged within 25 s of the BATSE trigger, of which two
pairs and a cluster of three arrive on a sufficiently short,
$\lesssim 0.2$ s timescale that deadtime effects could have
reduced the detected number of photons. Based on background
observed before the trigger time of the GRB, only 0.04 events are
expected by chance within 25 s \citep{som94}, so background should
not be important for this GRB. This GRB was very intense at BATSE
energies, but because most of this emission appeared in a first
very short, very intense pulse, the BATSE 20 keV -- 2 MeV fluence
is smallest of the 5 spark-chamber GRBs.

GRB 940217 \citep{hur94}, with spark-chamber emission extending to
$> 100$ min after the GRB trigger, has associated with it the
highest energy photon of all EGRET GRBs, $\approx 18$ GeV. A total
of 28 $\gamma$-ray photons were imaged by the EGRET spark chamber
from this GRB, with 10, 8, and 10 appearing during the $\approx
90$ s episode of strong BATSE emission, after the BATSE emission
and before Earth occultation occurring at $\approx 900$ s after
the GRB trigger, and after the GRB appears from Earth occultation
at $\sim 3700$ to 5400 s after the GRB.  For these three periods,
respectively, 0.39, 1.8, and 2.9 background $\gamma$ rays are
expected. Because the GRB was so extended in time and the $\gamma$
rays arrive one or more seconds apart, deadtime effects can be
considered to be small for this GRB.  The duration of Earth
occultation, representing $\gtrsim 62$\% of the known period of
activity of GRB 940217, therefore represents a major uncertainty
in the fluence measurement of this GRB.

Seven photons were imaged within 21 s of the GRB 940301 BATSE
trigger \citep{sch95}, though appearing sparsely distributed in
time so that deadtime effects are probably negligible. Background
is also likely to be small for this GRB, as no other $\gamma$ rays
are detected over a period of 150 s around the time of the GRB
except for the 7 photons mentioned.

For completeness, two other GRBs from the CGRO era should be
mentioned. This includes GRB 910814, with $\gamma$ rays observed
up to $\approx 50$ MeV with the TASC \citep[when the spark chamber
was disabled due to Earth occultation; in any case this event was
at 38.7$^\circ$ from the EGRET boresight;][]{kwo93}. The $>20$ keV
fluence of GRB 910814 was $\cong 2.51\times 10^{-4}$ ergs
cm$^{-2}$ \citep{mee96}. Late in the CGRO mission, when the EGRET
spark chamber gas was mostly depleted, GRB 990123 was observed
with COMPTEL to $\approx 18$ MeV. This was a bright GRB, with a
$>20$ keV fluence $\cong 5.1\times 10^{-4}$ ergs cm$^{-2}$
\citep{bri99}. At redshift $z = 1.6$ \citep{kul99}, it provided
the first confident lower limits on bulk Lorentz factor using
internal $\gamma\gamma$ opacity arguments \citep{bar06}. Joint
analysis of BATSE and EGRET-TASC  GRBs indicates excess, $\gtrsim
1$ MeV emission, though not necessarily extending to $\gtrsim 100$
MeV, in bright, long duration GRBs \citep{gon09}. The $>20$ keV
fluence of the most notable of these bursts, GRB 941017, was
$\cong 3\times 10^{-4}$ ergs cm$^{-2}$.

\begin{figure}[t]
\includegraphics[width=3.1in]{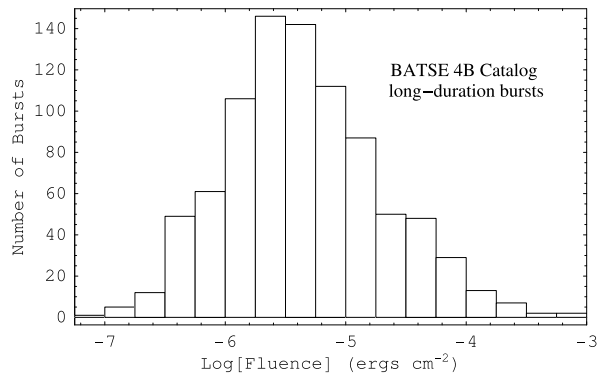}
\caption{Histogram of BATSE 4B catalog GRB fluence distribution;
data from \citet{pac99}. %
\label{fig2}} %
\end{figure}

Comparison of the distribution of the fluences of GRBs shown in
Fig.\ \ref{fig2} in the BATSE 4B catalog with the fluences of the
EGRET spark chamber GRBs in Table 2, and notable BATSE-TASC GRBs
in the preceding paragraph, shows that the high-energy emission
is, for the long duration GRBs, always associated with the most
fluent GRBs. This is characteristic of a sensitivity limited
detection of high-energy $\gamma$-ray emission. Consequently, we
can use the increased capabilities of the Fermi telescope with
respect to EGRET to make straightforward predictions by scaling
from the EGRET results. Note that the weakest GRB in Table 2 in
terms of fluence---GRB 930131---is also possibly a short, hard
GRB.

\subsection{Error from Finite Photon Number}

We consider a major error on fluence measurement to result from
statistical fluctuations due to the small number of photons imaged
by the EGRET spark chamber and associated with the GRB,
particularly since we limited the spark chamber events to those
with energy between 100 MeV and 5 GeV. It is a simple matter to
make this error estimate using a Monte Carlo simulation, and then
add other errors to get a complete error estimate.

Let the integrated GRB photon number spectrum be denoted by
$dN/dEdA$. The fluence measured between energies $E_1$ and $E_2$
is given by
\begin{equation}
F(E_1,E_2) = \int_{E_1}^{E_2} dE \; E \left({dN\over dE dA}\right)\;.
\label{FE1E2}
\end{equation}
The number of source photons is given by
\begin{equation}
N_s(E_1,E_2) = \int_{E_1}^{E_2} dE \; A(E,\theta) \left({dN\over dE dA}\right)\;.
\label{NsE1E2}
\end{equation}
The probability of detecting a photon of energy $E$ is simply
given by inverting the expression
\begin{equation}
r = {\int_{E}^{E_2} dE \; A(E,\theta) (dN/dE dA)\over \int_{E_1}^{E_2} dE \; A(E,\theta) (dN/dE dA)}\;,
\label{rinvert}
\end{equation}
for $E$, where $r$ is a random number uniformly distributed
between 0 and 1.

In the EGRET (and Fermi) energy range, $A(E,\theta)\cong const$
within a factor of $\approx 1.8$ for $100 < E({\rm MeV}) < 5000$
and $\theta < 30^\circ$, and can be  approximated by a power law
over this range, so we write $A(E,\theta) = A_0
E^{-u}$.\footnote{See Fig.\ \ref{fig1} and
www-glast.slac.stanford.edu/software/IS/glast\_\\
lat\_performance.htm; restricting the photon energies to $>200 $
MeV makes this an excellent approximation.} The effect of changing
$\theta$ on $A(E,\theta)$ over the duration of the GRB is
negligible for EGRET, which was a pointing telescope, but can be
important in Fermi observations of GRBs. More accurate numerical
calculations can be made following the analytic inversion given
here.

Assuming that $dN/dEdA \propto E^{-\alpha_1}$, the inversion of
eq.\ (\ref{rinvert}) is trivial, and we obtain for the detected
energy the value
\begin{equation}
E(r) = \left[ E_2^{1-s} +r(E_1^{1-s} - E_2^{1-s})\right]^{1/(1-s)}\;, %
\label{E(r)}
\end{equation}
where $s = u +\alpha_1$. The mean photon energy in the energy
range $E_1 \leq E \leq E_2$ is
\begin{equation}
\langle E\rangle  = \left( {E_1^{2-\alpha_1} - E_2^{2-\alpha_1}\over {E_1^{1-\alpha_1} - E_2^{1-\alpha_1}} }\right)
\;\left({\alpha_1 - 1\over \alpha_1-2}\right)\;.
\label{meanenergy}
\end{equation}
The error in fluence due to the finite number, $N_\gamma$, of
photons detected is then given by the root mean square deviation
of the detected photon energies from the average photon energy,
that is,
\begin{equation}
\Delta E = \sqrt{      { \sum_{i = 1}^{N_\gamma}[E(r_i) - \langle E\rangle]^2\over N_\gamma} }\;.
\label{fluence_error}
\end{equation}

\begin{figure}[t]
\includegraphics[width=3.50in]{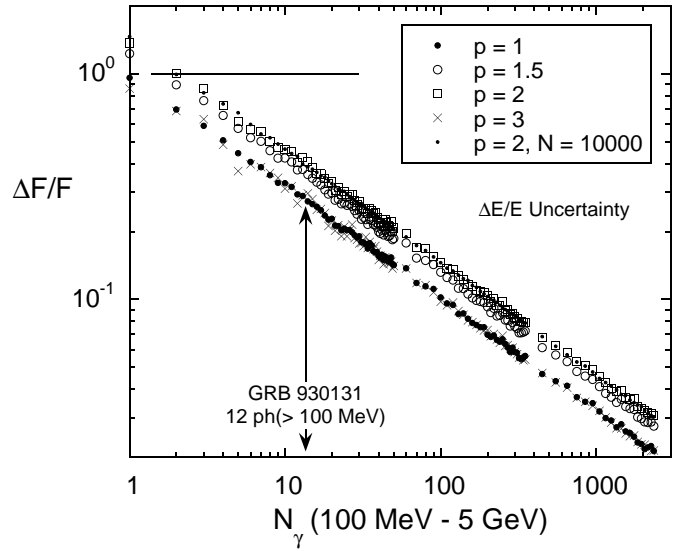}
\caption{Fractional uncertainty in measurement of the 100 MeV -- 5
GeV $\gamma$-ray fluence due to finite number of photons from the
GRB, for an underlying photon spectra described as a power-law
with index $\alpha_1 = p$. Calculation uses $N = 1000$ sets of
$N_\gamma$ photons chosen randomly, except where noted.%
\label{fig3new}}
\end{figure}

The fractional fluence error, $\Delta F/F$, as a function of the
number of detected photons $N_\gamma(100$ MeV -- 5 GeV) in  the
photon energy range between $E_1 = 100$ MeV to $E_2 = 5$ GeV range
is shown in Fig.\ \ref{fig3new} for the simplest approximation of
a flat effective area, so that $u = 0$. Because the fluence is
weighted by energy, and we have restricted the photon energy to a
finite range, a spectrum with $\alpha_1 = 2$ gives the largest
fluence uncertainty, which is well described by the
function

\begin{equation}
{\Delta F\over F} \cong {1.45\over \sqrt{N_\gamma}}\;,
\label{dFoverF}
\end{equation}
where that the factor of 1.45 is derived from fitting the data
(with p = 2) in Figure~\ref{fig3new}. We can take this expression
as providing the maximum error in fluence from the statistical
uncertainty due to the finite number of photons. If there is a
priori knowledge that the intrinsic spectrum has a specific number
index, e.g., $\alpha_1 = 2.2$, then the numerical coefficient of
eq.\ (17) becomes rather smaller, $\approx 1.4$. Eq.\ (17) is a
conservative estimate of fluence uncertainty, though it still
omits other possible systematic uncertainties. Note that in the
EGRET era, GRB 930131 had the largest number of photons,
$N_\gamma(100$ MeV -- 5 GeV)$\approx 12$, $\gamma$ rays, in the
0.1 -- 5 GeV range. As is apparent from Fig.\ \ref{fig3new},
finite photon number gives the major source of error to the
fluence measurement of EGRET spark-chamber GRB.

In comparison, the error associated with relative energy
uncertainty is, as already noted, at the $\simeq 10$\% level for
EGRET. For the Fermi Gamma ray Space Telescope, the relative
energy uncertainties at photon energies of 0.1, 1, and 10 GeV are
$\approx 17$\%, 9\%, and 8\% for normal incidence, and $\approx
15$\%, 9\%, and 5\% at 60$^\circ$ off-axis \citep{atw09}. Because
this can be a calibration uncertainty affecting all photon energy
measurements systematically rather than statistically, this
uncertainty on the fluence will start to dominate for those rare
GRBs detected with Fermi with more than $\sim 50$ photons between
100 MeV and 5 GeV. Effective area uncertainties also appear at the
$\sim 10$\% level. Consequently, the uncertainty from the finite
number of photons dominates fluence measurement for EGRET GRBs.

From the discussion of EGRET GRBs in Section 3.1 and their
properties in Table~\ref{tbl-2}, we can now reliably estimate the
fluence uncertainty associated with finite number of photons,
background and deadtime effects measured with EGRET in the 0.1 --
5 GeV energy range. For GRB 910503, we take $N_\gamma(100$ MeV --
5 GeV) = 2, essentially giving 100\% uncertainty in the
measurement of fluence (taking twice this value would give a
2$\sigma$ upper limit). For GRB 910601, we again conservatively
take $N_\gamma(100$ MeV -- 5 GeV) = 2 in view of possible
background. For GRB 930131, we take $N_\gamma(100$ MeV -- 5 GeV) =
12, keeping in mind that a deadtime correction in the absolute
fluence by tens of percent or more could be required, depending on
the distribution of $\gamma$-ray arrival times, which Fermi will
measure.  For GRB 940217, we take $N_\gamma(100$ MeV -- 5 GeV) =
8, subtracting two background photons that could have contributed
during the first prompt and early afterglow in this GRB (note that
no EGRET photons in the 100 MeV -- 5 GeV range were detected after
Earth occultation). For GRB 940301, we take $N_\gamma(100$ MeV --
5 GeV) = 5.

\subsection{Fluence Estimate from EGRET Photon Event List}

\begin{figure}[t]
\includegraphics[width=3.50in]{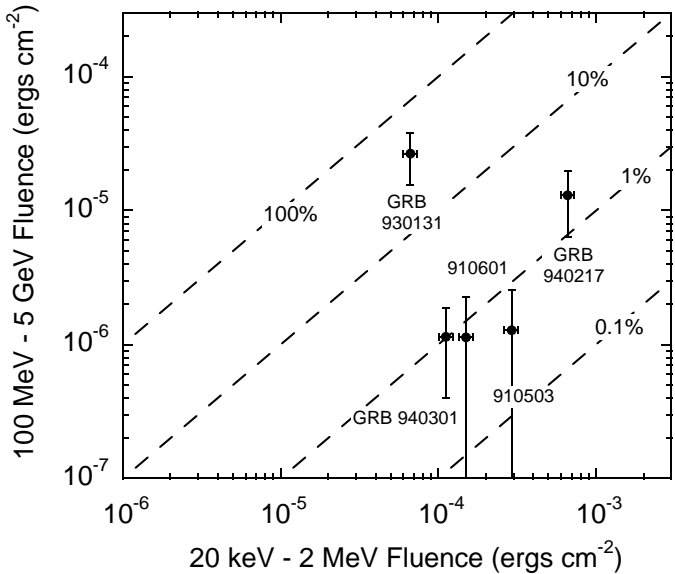}
\caption{Diagram plotting fluence measured between 100 MeV and 5 GeV with
the EGRET spark chamber vs the 20 keV -- 2 MeV fluence measured with BATSE
for the 5 EGRET spark chamber GRBs. The EGRET fluence for GRB 940217 could easily
be a factor of two larger due to Earth occultation during the active phase of this
GRB.
\label{fig4new}}
\end{figure}

The statistical errors associated with the finite number of
detected photons, being the largest source of error for the EGRET
GRBs, are used with the values in Table~\ref{tbl-2} and eq.\
(\ref{fluence}) to  produce Fig.\ \ref{fig4new}. This figure
displays the 100 MeV -- 5 GeV fluence measured with the EGRET
spark chamber vs. the $20$ keV -- 2 MeV fluence measured with
BATSE. Bolometric corrections to a larger energy range are usually
small if $E_{\rm pk}$, the peak photon energy of the $\nu F_\nu$
spectrum, falls in this waveband. This is not always the case for
these GRBs, and comparison of this figure with reported results in
the Fermi era must consider bolometric corrections.

This fluence-fluence diagram provides, potentially, one way to
discriminate between different classes of GRBs on the basis of
their $\gamma$-ray radiative efficiency. A much more accurate
comparison will be provided for those GRBs with known redshift.
Because deadtime and background effects have been corrected from
these EGRET spark chamber GRBs, the fluence-fluence diagram gives
a picture where the long-duration GRBs have a radiative efficiency
$\rho$ in  $>100$ MeV $\gamma$ rays  of $\approx 1$\% or a few \%
of the bolometric energy output. Assuming no class bias in the
values in Table~\ref{tbl-2}, these give the average deduced
radiative efficiency $\langle \rho \rangle \cong 9$\% (column 8).
This suggests that we use as a mean fluence ratio a number between
$\approx 10$, before considering broad ranges of value of $\rho$
or even separate classes with different $\rho$ values.

\begin{deluxetable*}{lcccc}
\hskip-1.in
\tabletypesize{\scriptsize} %
\tablecaption{Fermi and AGILE GRBs/yr FOV ($\approx
1/5^{th}$) Predictions %
\label{tbl-3}} %
\tablewidth{0pt} %
\tablehead{ %
\colhead{} & %
\multicolumn{2}{c}{spectral index of -2}    & %
\multicolumn{2}{c}{spectral index of -2.2}  \\ %
\colhead{Threshold Energy} & %
\colhead{$\rm Fermi $}     & %
\colhead{$\rm AGILE $}     & %
\colhead{$\rm Fermi $}     & %
\colhead{$\rm AGILE $}       %
} %
\startdata
$>$ 30 MeV \hskip+0.05in ($> 5$ photons) & %
30$^a$ (18)$^b$                          & %
7$^a$  (4)$^b$                                   & %
12$^c$ (6)$^d$                                   & %
2$^c$  ($<$ 1)$^d$                                 %
\\
$>$ 100 MeV ($> 5$ photons)          &  %
\hskip-0.05in 20 \hskip+0.07in (12)  &  %
\hskip+0.1in  2 \hskip+0.05in ($<$ 1)&  %
\hskip+0.05in 8 (4)                  &  %
\hskip-0.12in $<$ 1 ($<$ 1)             %
\\
$>$ 1 GeV \hskip+0.13in ($> 5$ photons) & %
\hskip+0.08in 4 \hskip+0.07in ($<$ 2)   & %
\hskip-0.12in $<$ 1 (0)                 & %
\hskip-0.08in $<$ 1 (0)                 & %
\hskip-0.23in $<$ 1 (0)                 %
\enddata
\tablecomments{This column has value with fluence ratio of
$^a$10\%; $^b$5\%; $^c$8\%; $^d$4\%.}

\end{deluxetable*}

\begin{figure*}[t]
\includegraphics[width=7.in]{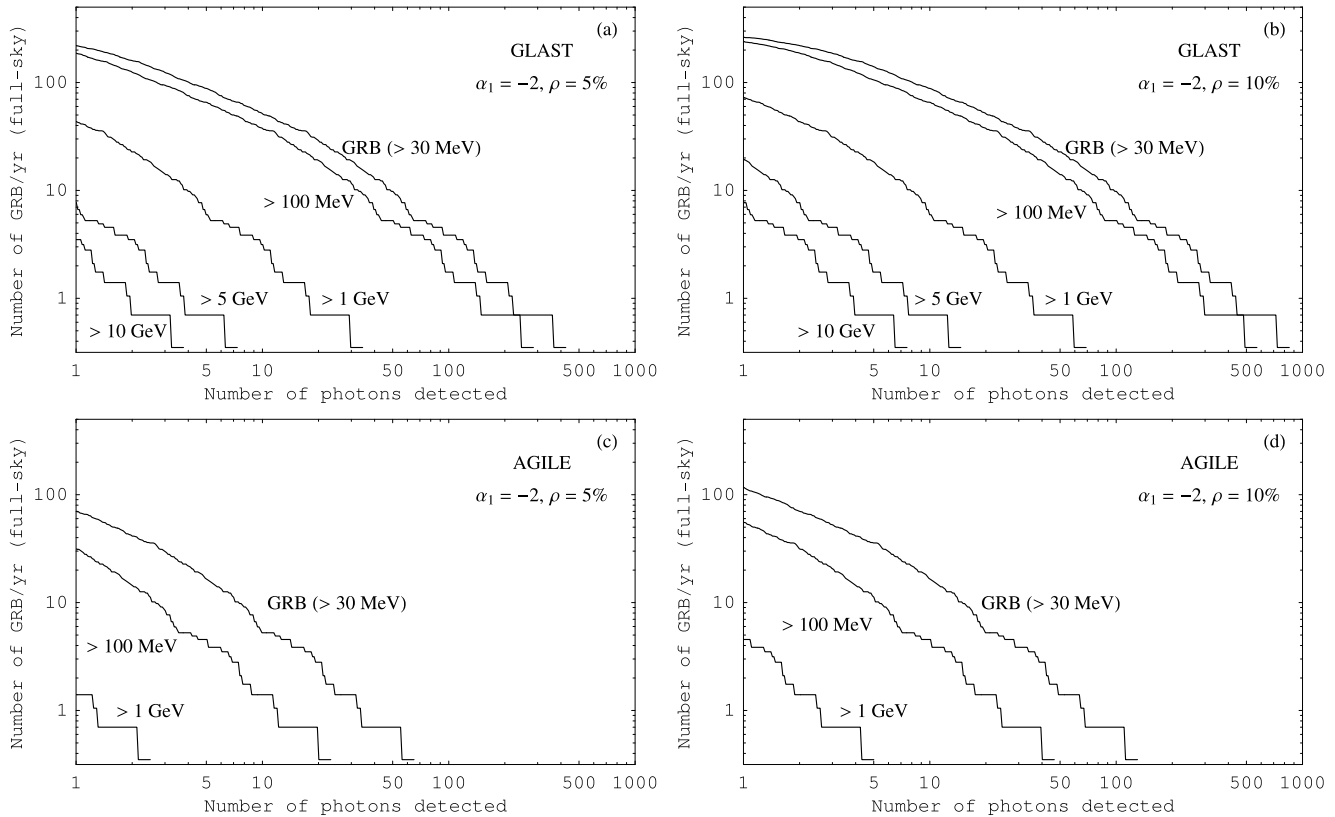}
\caption{Prediction for the number of long-duration GRBs per year
full-sky Fermi LAT (a and b) and AGILE (c and d) will see based on
a measured fluence ratio extrapolated from a $-2$ photon spectum
from EGRET/BATSE to EGRET LAT energies. The curves describe the
burst rate assuming a $-2$ photon spectra for different energy
thresholds.\\}
\label{f5}
\end{figure*}
\begin{figure*}[t]
\includegraphics[width=7.in]{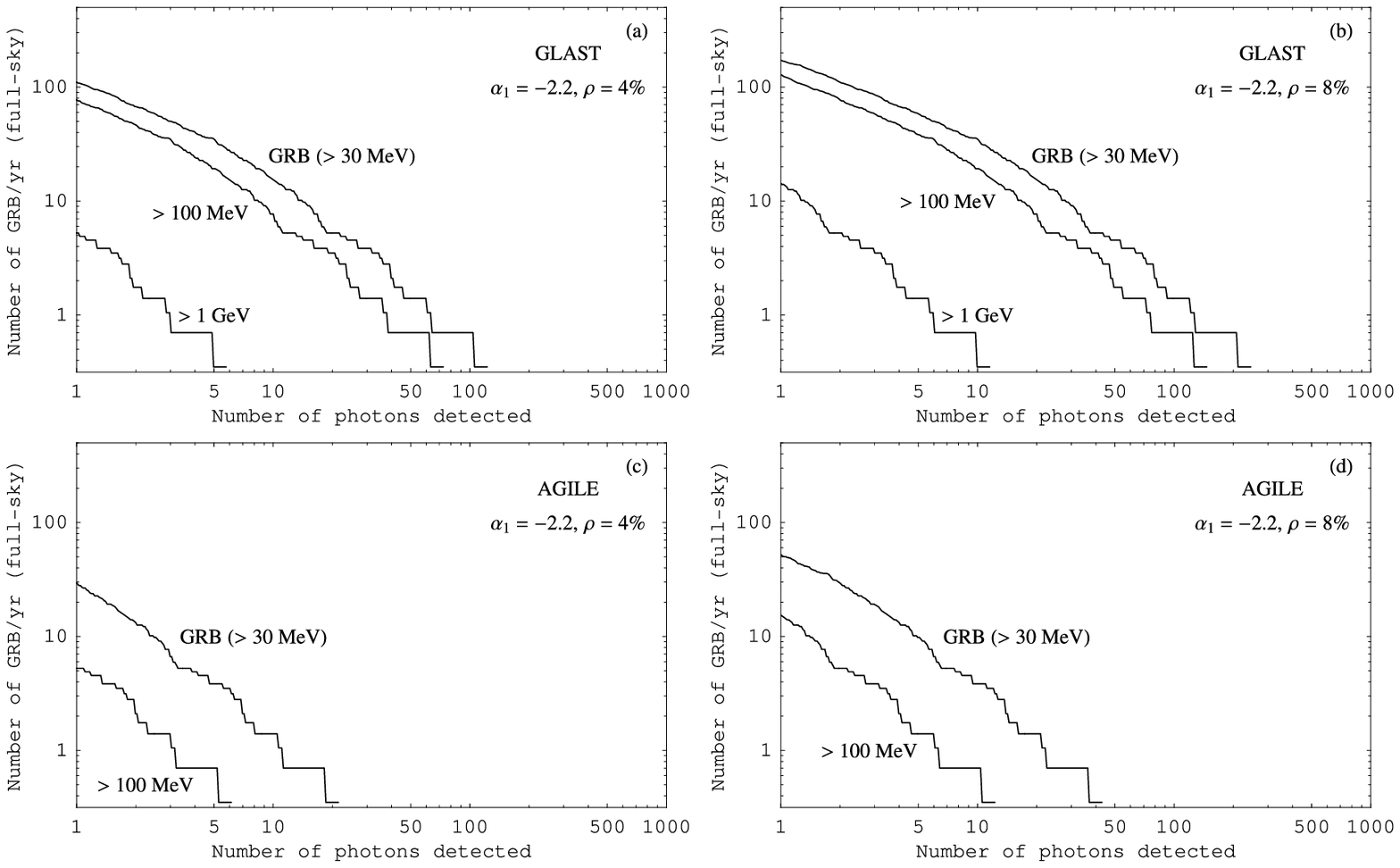}
\caption{Prediction for the total number of GRBs/yr full-sky that
Fermi LAT (a and b) and AGILE (c and d) will detect for GRB
spectra as described in Fig.~3 for an assuming $-2.2$ photon
spectra.\\}

\label{f6}
\end{figure*}

\section{Alternate Fluence Estimate}

As a check on $\rho$, we also make an independent estimate using
the number $N_\gamma(>100$ MeV) of $>100$ MeV $\gamma$-ray photons
that were detected by EGRET \citep{din95}, assuming a hard
component of the spectrum with  spectral indices $\alpha_1 = -2$
and $\alpha_1 = -2.2$. The EGRET fluence in the energy range of
100 MeV to 5 GeV can be expressed as
\begin{equation}
 \Phi_{\rm EGRET} = N_\gamma (> 100 {\rm~MeV})\;{
\int^{5 \,{\rm GeV}}_{100\, {\rm MeV}} E^{1-\alpha_1} dE \over
\int^{5\,{\rm GeV}}_{100\, {\rm MeV}} E^{-\alpha_1} A(E,\theta) dE
}\;,
\end{equation}
where $A(E,\theta)$ is the angle-dependent EGRET effective area
obtained by interpolating the data in Figure~14 of \citet{tho93}.
From Table~\ref{tbl-2} we find that the average measured fluence
ratios $\rho$ for the 5 spark chamber bursts are $\rho \approx
9\%$ (column 10) and $\rho \approx 7\%$ (column 12) for spectra
indices of $\alpha_1 = -2$ and $-2.2$, respectively. This is
consistent with our previous estimate of $\approx 9$\%, because
that referred to the $> 100$ MeV--5 GeV fluence to 20 keV--2 MeV
fluence ratio, rather than the $> 30$ MeV--5 GeV fluence to 20
keV--2 MeV fluence ratio, the latter of which is expected to be
$\approx 50$\% larger. However, these average fluence values of
9\% and 7\% are dominated by the boundary of long-soft and
short-hard GRB 930191 with the fluence ratios of 40\% and 30\% for
spectra indices of -2 and -2.2, respectively. Four other bursts
are long GRBs with the average fluence ratios of 1.3\% and 1\%.
Hence, we expect the average fluence for the long GRBs to be
larger, conservatively, we assume values of 5\% and 4\% for
spectra indices of $\alpha_1 = -2$ and $-2.2$, respectively, to
represent long-duration GRBs.

However, as previously noted, the total $\gamma$-ray flux is
probably larger than EGRET measured. The $\gamma$-ray emission
from GRB 940217 \citep{hur94} likely persisted during the period
of Earth occultation, easily increasing the number of photons by
$\sim 2$. Deadtime effects, as seen in GRB 930131 \citep{som94},
could have caused the number of $\gamma$ rays to be undercounted.
Besides, there is the example of GRBs with anomalous hard tails,
like GRB 990104 \citep{wbr02} and GRB 941017 \citep{gon03}, the
latter with fluence in the $\approx 1$ -- $100$ MeV range
comparable to or greater than the sub-MeV fluence. Further, there
is the low-significance Milagrito detection of the X-ray
flash-type GRB 970417a \citep{atk03}, which would not only have to
be at low redshift $z \lesssim 0.3$ but also have much greater TeV
than MeV fluence.  Hence we expect the average fluence ratios
$\rho$ between EGRET and BATSE to be $\approx 10$\% and $\approx
8$\% for spectra indices of $\alpha_1 = -2$ and $\alpha_1 = 2.2$,
respectively.

We use $\rho = 0.1$ for model with $\alpha_1 = -2$ and $\rho =
0.08$ for model with $\alpha_ 1 = -2.2$ to define the value of
$k_G$. The estimated total number of long-duration GRBs that get
detected by any detectors are based on the BATSE 4B fluence
distribution \citep{pac99}, shown in Fig.~\ref{fig2}, assuming an
annual GRB rate of 670 GRB/yr, including untriggered GRBs with
photon fluxes smaller than 0.3 cm$^{-2}$ s$^{-1}$ \citep{ban02}.
The BATSE 4B Catalog contains 1292 bursts total, including short
and long duration GRBs. There are a total of 872 long-duration
GRBs that are identifiable in the BATSE 4B Catalog. When we
estimate the number of source counts over one year period, the
ratio 872/1292 is applied to represent the long-duration GRB
predictions.

\section{Results}

In Figs.\ \ref{f5} and \ref{f6}, we plot full-sky Fermi LAT and
AGILE GRB detection rates, assuming that the Band-spectrum extends
to $\gtrsim 100 $ MeV energies. This simple model predicts that
$\approx 30, 20$, and $4$ GRBs will be observed per year with the
Fermi LAT with at least 5 photons with energies $E > 30$ MeV, $E >
100$ MeV, and $E> 1$ GeV, respectively. This is assuming a hard
spectral index  $\alpha_1  = -2$ and a fluence ratio $\rho =
10$\%. For $\alpha_1  = -2.2$, the model predicts that $\approx
12$, $\approx 8$, and probably no GRBs per year should be detected
with at least 5 photons with $E > 30$ MeV, $E > 100$ MeV, and $E>
1$ GeV, respectively, now assuming that $\rho =  8$\%.

The analysis also predicts that $\approx 7$ and $\approx 2$ GRBs
per year with at least 5 photon counts at $E >$ 30 MeV and $E >$
100 MeV, respectively, should be detected by AGILE assuming
$\alpha_1  = -2$ and $\rho =10$\% (see Fig.~5d). For $\alpha_1  =
-2.2$, the model predicts that $\approx 2$ and probably no GRBs
with at least 5 photons with $E > 30$ MeV and $E > 100$ MeV,
respectively, will be detected by AGILE per year with $\rho = 8$\%
(see Fig. 6d). A consistently smaller number of GRBs/yr full-sky
is predicted for both Fermi and AGILE for the steeper spectral
indices, as expected. In Table~\ref{tbl-3}, we summarize the
predicted number of GRBs per yr with at least 5 photons greater
than a specific $\gamma$-ray photon energy in the FoV, $\approx
2.5$ sr, of the Fermi LAT and  AGILE, considering GRBs with hard
spectral indices $\alpha_1 = -2$ and $\alpha_1 = -2.2$, and for
fluence ratios $\rho = 10$\%, 8\%, 5\%, and 4\%.

These models assume that all GRBs have power law with $\alpha_1
\approx -2$ and 10\% fluence ratio or $-2.2$ and 8\% fluence
ratios, and these assumptions would be first to be examined if
Fermi LAT shows strong violations of these predictions. For
$\alpha_1 = -2$ and $\rho = 10$\%, we estimate from Fig.\ 5 that
$\approx 1$ -- 2 GRBs per year will be detected with the Fermi
telescope that have more than 100 $\gamma$ rays with $E\gtrsim
100$ MeV. The models also predicts that the GBM fluences of
long-duration GRBs detected with the LAT should be large, $\gtrsim
10^{-5}$ ergs cm$^{-2}$. The distribution of fluences can be
compared with predicted distributions for Swift and the Fermi GBM
\citep{ld07}.

From this treatment of the fluence ratio, we can estimate the
diffuse extragalactic $\gamma$-ray background produced by GRBs
using the 4B BATSE fluence catalog based on the EGRET/BATSE
fluence ratio. From the 4B Catalog there are a total of 1292 GRBs,
of which 872 bursts are long-duration GRBs. The total average
fluence for the long-duration GRB is about $\approx 1.40 \times
10^{-2} \rm ~ergs~cm^{-2}$. Assuming 670 GRBs/yr, we estimate that
less than 1\% of the diffuse extragalactic $\gamma$-ray background
could come from GRBs assuming $\approx 8\%$ or 10\% fluence ratio.
Unless there is a large class of $\rho\approx 1$ GRBs, which were
not found with the EGRET spark chamber, this suggests that GRBs
gives very little contribution to the diffuse extragalactic
$\gamma$-ray background \citep[e.g.,][]{der06,cdz07}.

\section{Conclusions}

As we enter the Fermi era, it is worthwhile to make final
predictions based on results from the EGRET CGRO era. Here we have
adopted an approach making use of mean fluence ratios in the EGRET
100 MeV -- 5 GeV and BATSE 20 keV -- 2 MeV bandpasses. Obviously
this already introduces significant K-corrections when calculating
the intrinsic properties of the GRB sources; a sample of GRBs with
known redshifts is clearly preferred.

Nevertheless, we fix the fluence ratios in these bands to obtain
straightforward predictions for all-sky detection rates with
Fermi, using on-axis effective areas and the BATSE 4B fluence
distribution.  Multiplication by $1/5^{\rm th}$ gives the final
detection rate with the Fermi LAT tracker and calorimeter
\citep{atw09}

For a 10\% fluence ratio and a hard component with $-2$ index, we
estimate $\approx 1$ -- 2 per month with more than 5 photons with
$E> 100$ MeV $\gamma$ rays in the LAT FoV, and 1 or 2 per year
with more than 100 $\gamma$ rays with $E > 100$ MeV. A bright GRB
for spectral modeling, with more than $\approx 100$ photons with
$E>100$ MeV, is predicted once or twice per year. We note that
with the use of $-2$ fluence spectrum, the detection of a $> 10$
GeV photon from a GRB is considered very improbable, barring the
existence of an additional component harder than $-2$.

Furthermore, as the Fermi GBM and the Fermi LAT instruments begin
to collect data from GRBs, we recommend to plot LAT fluence vs.\
GBM fluence and, more importantly, mean LAT GRB spectral index
vs.\ LAT/GBM fluence ratios $\rho$ to search for distinct classes
of GRBs distinguished by their high-energy properties.

\acknowledgements We thank David Band, Michael Briggs, Valerie
Connaughton, Brenda Dingus, Magda Gonz\'alez, Chryssa Kouveliotou,
Julie McEnery and Nicola Omodei for many interesting discussions
and useful suggestions, and especially to Nicola Omodei and Julie
MeEnery for a really difficult and constructive report. We also
thank the referee for comments, corrections, and useful
suggestions. The work of CD is supported by the Office of Naval
Research and NASA Fermi Science Investigation DPR-S-1563-Y, which
also supported the research of TL at the Naval Research
Laboratory.

{}

\end{document}